# Fixed-Gain Augmented-State Tracking-Filters

Hugh L. Kennedy

DST Group, NSI Division, Edinburgh, SA 5111 Australia

(e-mail: Hugh.Kennedy@dst.defence.gov.au)

*Abstract*—A procedure for the design of fixed-gain tracking filters, using an augmented-state observer with signal and interference subspaces, is proposed. The signal subspace incorporates an integrating Newtonian model and a second-order maneuver model that is matched to a sustained constant-g turn; the deterministic interference model creates a Nyquist null for smoother track estimates. The selected models provide a simple means of shaping and analyzing the (transient and steady-state) response of tracking-filters of elevated order.

I. INTRODUCTION

The exploitation of prior knowledge by the Kalman filter (KF) gives rise to a variable gain and adaptive behavior, which is highly desirable in automatic target-tracking systems. Knowing the parameters of measurement-noise and process-noise statistics, and the covariance of the state estimate that follows, provides additional leverage that may be used to: accelerate the removal of bias errors during track establishment; handle changing sensor/target characteristics; accommodate variable revisit intervals; and manage measurement-to-track assignment ambiguities that arise in the presence of clutter, multiple closely-spaced targets, and missing measurements. When this flexibility is not required, maintaining the covariance matrix and evaluating the resulting filter gain on each update is an expensive and unnecessary overhead [1]; furthermore, worst-case tracking accuracy may be severely degraded if the assumed noise parameters are very wrong.

Interest in simple and robust tracking filters for use in embedded or mission-critical systems, massively parallel implementations, and diverse applications – i.e. outside of the traditional aerospace domain – involving signals that exhibit low-frequency trends [2],[3], such as computer-vision, robotics, communications, and speech-processing [4]-[6], continues to grow. However, due to the remarkable success of the KF in these areas over the last 50 years, there are unfortunately very few fixed-gain alternatives.

The most common approach begins with a KF for position-velocity(-acceleration) states; the steady-state (SS) Kalman gain vector is then determined via the Riccati equations. For a KF of order $K$, with noise matrices $\boldsymbol{Q}$ and $\boldsymbol{R}$ of a specified form, this method yields closed-form expressions for $\alpha - \beta$ ($K = 2$) or $\alpha - \beta - \gamma$ ($K = 3$) filter coefficients, as a function of the dimensionless $\lambda_{SS-KF} = T_s^2 \sigma_Q / \sigma_R$ quantity, which combines the sampling period $T_s$ and noise parameters, where $\sigma_Q$ & $\sigma_R$ are the standard deviations of the process-noise and measurement-noise distributions, respectively [1]. This parameter is commonly referred to as the tracking index [7]. General procedures for integrating models of arbitrary order, for a wider filter bandwidth, are available [1],[8]; however, this is likely to degrade tracking performance without additional measures to attenuate noise and interference. The design of $\alpha - \beta(-\gamma)$ trackers for processing position and velocity measurements [9], and for handling nonlinearities in polar coordinate systems, is also being investigated [10].

An alternative class of methods begins by specifying some aspect of the transient response (for maneuver handling) and/or the SS response (e.g. signal bandwidth or noise gain), then proceeds with the discrete-time transfer-function/frequency-response of the tracking filter as the design object, rather than a set of assumed statistical distributions [11]-[20]. From these requirements, the $\alpha - \beta(-\gamma)$ parameters of a fixed-gain filter [12]-[18], or the process-noise parameter of a variable-gain KF are then derived [19],[20].

The merit of using deterministic analysis to design stochastic (i.e. Kalman) filters is questioned in [16] and the possibility of using state observers (e.g. Luenberger) with arbitrarily placed poles, is mooted; however, details of how such a design procedure might be adapted for this specific purpose are not provided. Posing the tracking problem in the transform domain (i.e. $z$ or $\omega$) simplifies the algebraic manipulations required to derive tracking filter coefficients [21],[22] and performance metrics [18]. It also opens the door to other design and analysis possibilities that have not hitherto been utilized in target-tracking problems.

Discrete-time shaping-filters with augmented states and a variable gain were originally formulated for handling colored noise in communications systems [23],[24]; they have since been extended to incorporate online system identification and applied to a variety of problems, such as speech processing [4]-[6]. In this paper, a steady-state variant is used to solve the Newtonian target-tracking problem: A deterministic process model – with signal and interference subspaces – and without random-noise terms is utilized, to provide a way of balancing conflicting bandwidth/noise-gain requirements in a filter of elevated order.

Unlike conventional methodologies described in the literature, that are based on the parameters of random process- and measurement-noise statistics, the tracker design and analysis procedure presented in this paper revolves around the tracking filter's frequency response. It is shown that deterministic process models may be crafted to shape the response, for improved maneuver handling at specified rates of turn and/or increased track smoothness at steady state. Pole-placement – standard in controller design but novel in tracker design – is then used to determine the convergence behavior after track initiation and input discontinuities (e.g. steps or ramps), i.e. the transient response.

The problem of tracking a point target in a Cartesian coordinate system is considered here – a distant target in an imaging (electro-optic or infrared) sensor, for example. Target and measurement dynamics are assumed to be approximately linear over the spatial and temporal scales of the sensor and the states are assumed to be independent and separable in both Cartesian dimensions. The derivation that follows deals with a single scalar measurement from one of the dimensions and measurement-to-track assignment is unambiguous – due to image feature descriptors, for instance. A constant time interval ($T_s$) between measurements is assumed.

## II. FORMULATION

The sequence of measurements collected by the sensor is assumed to be generated by an endogenous linear-time-invariant (LTI) process with signal and interference subspaces. The signal subspace is further assumed to contain target and maneuver terms. The target model governs baseline dynamics; it is an integrating process, with $K_{\text{tgt}}$ repeated poles at the origin of the complex $s$-plane. This basic Newtonian process may be augmented to include a complex conjugate pair of poles on the imaginary axis, near the origin, to model a low-frequency sinusoidal oscillation (i.e. a weave maneuver [11]), or a circular orbit (i.e. a sustained constant-g turn) in two Cartesian dimensions. Similar considerations have been used elsewhere to determine the bandwidth requirements for $\alpha - \beta(-\gamma)$ filters [17],[18]; however, explicitly including this term in the process model eliminates SS position tracking errors, for a perfectly matched maneuver. An interference model may also be included to remove unwanted high-frequency noise, or simply for the aesthetic of smoother tracks.

This fully deterministic formulation greatly simplifies the filter derivation and realization because the Riccati equations and covariance matrix manipulations are not required; moreover, despite the omission of explicit random-noise terms, the resulting fixed-gain filters confer excellent high-frequency noise suppression, via the interference model.

The "process" is a continuous-time natural system; whereas the "observer" is a discrete-time synthetic system, employing a model of the natural system, which is realized on a digital computer. The two systems are connected via a one-way flow of sampled sensor measurements. Feedback exists within both systems but feedback between the two systems is not considered here. In this section it is assumed that the process model is perfect; the impact of modelling errors is discussed in the section that follows. To assist the reader, a list of all symbols used in this paper is provided in Appendix C.

A. Target process model

The linear state-space (LSS) representation, of this continuous-time process, in controller-canonical form (CCF), is as follows:

$$\dot{\boldsymbol{w}}_{\text{tgt}}(t) = \boldsymbol{A}_{\text{tgt}}\boldsymbol{w}_{\text{tgt}}(t), y_{\text{tgt}}(t) = \boldsymbol{C}_{\text{tgt}}\boldsymbol{w}_{\text{tgt}}(t) \text{ with}$$

$$\boldsymbol{A}_{\text{tgt}} = \begin{bmatrix} \boldsymbol{0}_{(K_{\text{tgt}}-1)\times 1} & \boldsymbol{I}_{(K_{\text{tgt}}-1)} \\ 0 & \boldsymbol{0}_{1\times(K_{\text{tgt}}-1)} \end{bmatrix}_{K_{\text{tgt}}\times K_{\text{tgt}}} \text{ and}$$

$$\boldsymbol{C}_{\text{tgt}} = \begin{bmatrix} 1 & \boldsymbol{0}_{1\times(K_{\text{tgt}}-1)} \end{bmatrix}_{1\times K_{\text{tgt}}} \quad (1)$$

where $y_{\text{tgt}}(t)$ and $\boldsymbol{w}_{\text{tgt}}(t)$ are the output signal and the process states; furthermore, $\boldsymbol{I}_N$ is an $N \times N$ identity matrix and $\boldsymbol{0}_{M\times N}$ is an $M \times N$ matrix of zeros; thus $\boldsymbol{A}_{\text{tgt}}$ is a zero matrix with ones along the 1st upper diagonal. The corresponding discrete-time model, for a constant sampling period of $T_s$, is found via the $s$ domain in the usual way [25],[26], using

$$\boldsymbol{G}_{\text{tgt}} = \mathcal{L}^{-1}\{\boldsymbol{\Phi}_{\text{tgt}}(s)\}\big|_{t=T_s} \text{ where} \quad (2a)$$

$$\boldsymbol{\Phi}_{\text{tgt}}(s) = \left(s\boldsymbol{I}_{K_{\text{tgt}}} - \boldsymbol{A}_{\text{tgt}}\right)^{-1} \text{ and} \quad (2b)$$

$\mathcal{L}^{-1}$ is the inverse Laplace transform ($t \leftarrow s$). The state-transition matrix $\boldsymbol{G}_{\text{tgt}}$, is an upper triangular Toeplitz matrix with the elements along the $k$th off-diagonal (for $k = 0 \ldots K_{\text{tgt}} - 1$) equal to $\mathcal{G}(k;T_s) = T_s^k/k!$. As the diagonal elements of this triangular matrix are equal to unity, the discrete-time model of the target signal has $K_{\text{tgt}}$ repeated poles in the complex $z$-plane at $z = 1$, for a singularity at dc, i.e. at $\omega = 0$, where $\omega$ is the angular frequency (radians per sample).

B. Maneuver process model

This second-order term is defined in CCF as follows:

$$\dot{\boldsymbol{w}}_{\text{man}}(t) = \boldsymbol{A}_{\text{man}}\boldsymbol{w}_{\text{man}}(t), y_{\text{man}}(t) = \boldsymbol{C}_{\text{man}}\boldsymbol{w}_{\text{man}}(t) \text{ with}$$

$$\boldsymbol{A}_{\text{man}} = \begin{bmatrix} 0 & 1 \\ -\Omega^2 & 0 \end{bmatrix} \text{ and } \boldsymbol{C}_{\text{man}} = \begin{bmatrix} 1 & 0 \end{bmatrix} \quad (3a)$$

where $\Omega$ is the maneuver turn rate (rad/s) for a constant-speed orbit of radius $R$; thus

$$\boldsymbol{G}_{\text{man}} = \begin{bmatrix} \cos(\Omega T_s) & \sin(\Omega T_s)/\Omega \\ -\Omega\sin(\Omega T_s) & \cos(\Omega T_s) \end{bmatrix} \qquad (3b)$$

which has poles in the $z$-plane at $z = e^{\pm i\Omega T_s}$ where $i^2 = -1$. Note that $\boldsymbol{A}_{\text{man}}$ and $\boldsymbol{G}_{\text{man}}$ are for the continuous-time and discrete-time processes, respectively. The latter representation is derived from the former via the same standard procedure used to discretize the target process model in (2).

A second-order *damped* oscillator is used in [27] to model "wind sway or platform roll". It is an extension of the Singer model, which shifts a pole at the origin of the $s$-plane, left along the real axis [22],[28]. Although, this result is derived via the introduction of first-order auto-correlation in the noise input, its affect may be physically interpreted as a drag force acting on the target, for a second-order process, with position and velocity states, an acceleration input and a position output.

The *undamped* one-dimensional (1-D) oscillator used here in (3a), with poles on the imaginary axis of the $s$-plane at $\pm\Omega$, may seem somewhat unrealistic for a target's motion in 1-D; however, when two such models are placed in parallel, to represent independent motion in two orthogonal Cartesian axes, they may be used to perfectly model circular motion, when their internal states fall into antiphase. Unlike the 2-D coordinated turn models discussed in [29] & [30], the relative phase and magnitude of these parallel maneuver state-estimates in the observers that follow, are unconstrained and free to drift in a way that best accounts for the input data sequence; thus additional maneuver detection and model switching or mixing is not required [29],[30]. In addition to reducing computational complexity, this simplification allows linear–systems theory to be used for the design and analysis of the fixed-gain tracking filters that result. The cost of the proposed maneuver model is an increase in the measurement-noise gain, particularly if fast and tight turns are accommodated; however, this is partially offset by the interference process model.

C. *Interference process model*

Proceeding along similar lines, the interference process is a $K_{\text{int}}$th-order system with $K_{\text{int}}$ repeated poles at $z = -1$, for a singularity at $\omega = \pi$. In this case $\boldsymbol{G}_{\text{int}}$, which has the same Toeplitz structure as $\boldsymbol{G}_{\text{tgt}}$, is populated using $-\mathcal{G}(k; T_s)$ for $k = 0 \ldots K_{\text{int}} - 1$, thus $\boldsymbol{G}_{\text{int}} = -\boldsymbol{G}_{\text{tgt}}$ and $\boldsymbol{C}_{\text{int}} = \boldsymbol{C}_{\text{tgt}}$ when $K_{\text{int}} = K_{\text{tgt}}$.

The form of this model is not physically motivated; rather, it is adopted to attenuate very high-frequency noise, which is assumed to be particularly undesirable, from sources inside or outside the sensor. An interferer of any frequency may in principle be defined; however, using an oscillation frequency of $\pi$ allows real poles to be used, instead of complex conjugate pairs, for a lower filter order. Using a frequency at the extremum of the spectrum also minimizes distortion in the signal band near dc. A single pole is generally sufficient for this model; however, degenerate poles may be used to extend its influence to lower frequencies.

*D. Process model*

The discrete-time model of the process that generates the sampled measurement sequence (i.e. "the process") may now be defined as

$$\boldsymbol{w}_{\text{prc}}(n) = \boldsymbol{G}_{\text{prc}} \boldsymbol{w}_{\text{prc}}(n-1) \text{ and}$$

$$y_{\text{prc}}(n) = \boldsymbol{C}_{\text{prc}} \boldsymbol{w}_{\text{prc}}(n) \text{ where}$$

$$\boldsymbol{w}_{\text{prc}}(n) = \begin{bmatrix} \boldsymbol{w}_{\text{tgt}}(n) \\ \boldsymbol{w}_{\text{man}}(n) \\ \boldsymbol{w}_{\text{int}}(n) \end{bmatrix}_{K \times 1}$$

$$\boldsymbol{G}_{\text{prc}} = \begin{bmatrix} \boldsymbol{G}_{\text{tgt}} & \boldsymbol{0}_{K_{\text{tgt}} \times K_{\text{man}}} & \boldsymbol{0}_{K_{\text{tgt}} \times K_{\text{int}}} \\ \boldsymbol{0}_{K_{\text{man}} \times K_{\text{tgt}}} & \boldsymbol{G}_{\text{man}} & \boldsymbol{0}_{K_{\text{man}} \times K_{\text{int}}} \\ \boldsymbol{0}_{K_{\text{int}} \times K_{\text{tgt}}} & \boldsymbol{0}_{K_{\text{int}} \times K_{\text{man}}} & \boldsymbol{G}_{\text{int}} \end{bmatrix}_{K \times K}$$

$$\boldsymbol{C}_{\text{prc}} = [\boldsymbol{C}_{\text{tgt}} \quad \boldsymbol{C}_{\text{man}} \quad \boldsymbol{C}_{\text{int}}]_{1 \times K} \text{ and}$$

$$K = K_{\text{tgt}} + 2K_{\text{man}} + K_{\text{int}}, \text{ with } K_{\text{man}} \leq 1$$

$([\cdot]^T$ is the transpose operator). (4)

Note that the signal and interference models used here are just examples of possible processes that may be relevant in a tracking problem. Other process definitions and permutations are possible; however, this particular combination was found to provide sufficient flexibility and control over tracking behavior.

*E. Observer design*

As the process has $K$ poles on the unit circle, it is marginally stable. We now seek a stable observer, i.e. all poles inside the unit circle, to estimate the states of the augmented-state vector, placed in series with the process, that has the following discrete-time LSS representation:

$$\widehat{\boldsymbol{w}}_{\text{prc}}(n) = \boldsymbol{G}_{\text{prc}} \widehat{\boldsymbol{w}}_{\text{prc}}(n-1) + \boldsymbol{\mathcal{K}}\{x(n) - \hat{x}(n)\} \text{ or}$$

$w_{\text{obs}}(n) = G_{\text{obs}}w_{\text{obs}}(n-1) + H_{\text{obs}}x(n)$ and

$y(n) = C_{\text{obs}}w_{\text{obs}}(n)$ where: (5a)

$y(n)$ is the smoothed output of the observer

$C_{\text{obs}} = [C_{\text{tgt}}G_{\text{tgt}}(q) \quad C_{\text{man}}G_{\text{man}}(q) \quad 0_{1 \times K_{\text{int}}}]_{1 \times K}$

$w_{\text{obs}} = \hat{w}_{\text{prc}}, H_{\text{obs}} = \mathcal{K}, G_{\text{obs}} = G_{\text{prc}} - \mathcal{K}C_{\text{prd}}$

$C_{\text{prd}} = C_{\text{prc}}G_{\text{prc}}$ (i.e. a one-step-ahead predictor)

$\hat{w}_{\text{prc}}(n)$ is the estimate of $w_{\text{prc}}(n)$

$\mathcal{K}$ is the $K \times 1$ observer gain vector

$\hat{x}(n) = C_{\text{prd}}\hat{w}_{\text{prc}}(n-1)$ is the predicted input

$x(n)$ is the observer input, with $x(n) = y_{\text{prc}}(n)$. (5b)

In the above definitions, $q$ is the delay parameter (an integer, in samples) and $G_{\text{sig}}(q)$ is a signal state-transition matrix, for a time-step of $-qT_s$ seconds, i.e.

$$G_{\text{sig}}(q) = \Phi_{\text{sig}}(t)\big|_{t=-qT_s} = \{G_{\text{sig}}^{-1}\}^q = G_{\text{sig}}^{-q}. \quad (6)$$

Note that $C_{\text{prd}}$ (for $\hat{x}$) involves the signal and noise subspaces, whereas $C_{\text{obs}}$ (for $y_{\text{obs}}$) considers only the signal subspace.

The gain vector $\mathcal{K}$, is found in the usual way, via a coordinate transform, which reduces the process equations to a canonical form for the observable $\langle C_{\text{prd}}, G_{\text{prc}} \rangle$ pair [26]. The transform required is readily found using the observability matrices of both coordinate systems (see Appendix A for details); alternatively, the discrete-time versions of the Ackermann or the Bass-Gura formulae may be used to find the gain vector directly [25],[26].

The elements of the gain vector $\mathcal{K}$ are chosen to arbitrarily place the poles of the observer in the complex $z$-plane for the desired convergence characteristics. Let the $k$th pole of the observer be $\lambda_k$ (for $0 \leq k < K$). All poles must be placed in a way that results in a stable feedback observer (i.e. $|\lambda_k| < 1$). It is also desirable to have a system that is not too oscillatory (i.e. $\angle \lambda_k \approx 0$), with a rate of decay that is slow enough to attenuate white noise (i.e. $|\lambda_k| \to 1$), yet fast enough to quickly remove bias arising from discontinuities in the signal state and from system modelling errors (i.e. $|\lambda_k| \to 0$). Repeated real poles, i.e. $\lambda_k = p$, for all $k$, with $0 \leq p < 1$, are sufficient for low-frequency signals with a narrow bandwidth, or for signals that are sampled at a sufficiently high rate, and they are used exclusively here. However, a complex pair of poles may be required in very-wideband filters. Using $p = 0$ yields a so-called "deadbeat" observer, with an FIR. For a complex argument (arg), |arg| and ∠arg are the magnitude and angle operators, respectively.

*F. Filter design*

The state observer accepts a scalar-valued input $x(n)$, and produces a vector-valued output $\boldsymbol{w}_{\text{obs}}(n)$. In prediction, filtering, and smoothing, problems (using $q < 0$, $q = 0$ and $q > 0$, respectively), only the position element ($k = 0$) of the state, or its $D$th time derivative, are of interest. To analyze the response of this single-input/single-output (SISO) system, and to yield a minimal-complexity realization of the discrete-time transfer function $\mathcal{H}(z)$, it is convenient to determine a new coordinate transform that reduces (5) to a canonical form (see Appendix A for details) [25],[26],[31]. The $\boldsymbol{b}$ and $\boldsymbol{a}$ coefficients of the canonical filter realization of $\mathcal{H}(z)$, where $\mathcal{H}(z) = \mathcal{B}(z)/\mathcal{A}(z)$ and $\mathcal{A}(z) = (z - p)^K$, are then readily extracted from the observer equations in canonical form, such that

$$y(n) = \sum_{k=0}^{K-1} \boldsymbol{b}(k) x(n-k) - \sum_{k=1}^{K} \boldsymbol{a}(k) y(n-k). \quad (7)$$

SISO filters that compute the $D$th derivative of the output signal are found by operating on the state vector using an alternative measurement operator in (5b), which is used to select the desired state element for smoothing and analysis:

$$\boldsymbol{C}_{\text{obs}} = \begin{bmatrix} \boldsymbol{C}_{\text{tgt}} \boldsymbol{G}_{\text{tgt}}^{-q} \boldsymbol{\mathcal{D}}_{\text{tgt}} & \boldsymbol{C}_{\text{man}} \boldsymbol{G}_{\text{man}}^{-q} \boldsymbol{\mathcal{D}}_{\text{man}} & \boldsymbol{0}_{1 \times K_{\text{int}}} \end{bmatrix}_{1 \times K} \quad (8)$$

where $\boldsymbol{\mathcal{D}}_{\text{tgt}} = \boldsymbol{A}_{\text{tgt}}^D$ and $\boldsymbol{\mathcal{D}}_{\text{man}} = \boldsymbol{A}_{\text{man}}^D$, i.e. $D$ consecutive applications of the continuous-time state matrix.

The use of this observer results in a SISO filter with a "flatness" order of $L_{\text{dc}}$, $L_{\text{wb}}$ and $L_{\text{pi}}$, at the critical frequencies $\omega_c = 0$ (dc), $\omega_c = \pm \Omega T_s$ (wideband) and $\omega_c = \pi$ (Nyquist) respectively [2],[32], where $L_{\text{dc}} = K_{\text{tgt}}$, $L_{\text{wb}} = K_{\text{man}}$ and $L_{\text{pi}} = K_{\text{int}}$, for the complex frequency response $H(\omega) = \mathcal{H}(z)|_{z=e^{i\omega}}$, such that

$$\left. \frac{d^l}{d\omega^l} H(\omega) \right|_{\omega=\omega_c} = \rho_l(\omega_c), \text{ for } 0 \leq l < L;$$

where $\rho_l(\omega_c) = \left\{ \frac{d^l}{d\omega^l} \left( \frac{i\omega}{T_s} \right)^D e^{-iq\omega} \right\} \bigg|_{\omega=\omega_c}$ for

$\omega_c = 0$ and $\omega_c = \pm \Omega T_s$ (passband); and

where $\rho_l(\omega_c) = 0$ for $\omega_c = \pi$ (stopband). $\quad (9)$

For example, a fixed-lag smoother, with $D = 0$ and a lag of $q$ samples, configured using $K_{\text{tgt}} = K_{\text{int}} = L$ and $K_{\text{man}} = 0$, results in $\rho_l(0) = (-iq)^l$ and $\rho_l(\pi) = 0$, for $0 \leq l < L$. Flatness ensures that the filter behaves exactly/approximately as required at/near the critical design frequencies [2],[32]. Satisfaction of the dc constraints in (9) guarantees unbiased estimation of the kinematic target states, at steady state, in the absence of target maneuvers. The relationship between dc derivatives in the frequency domain and polynomial coefficients in the time domain (i.e. "vanishing-moment" theorem) is also used to design wavelets in [37].

### G. Tracker analysis

Bespoke closed-form expressions for the white-noise gain (WNG) of $\alpha-\beta(-\gamma)$ filters are available for various filter configurations [12]-[18]; however, for general cases (i.e. arbitrary order and lead/lag configurations), it may be simply evaluated for any stable transfer function $\mathcal{H}(z)$, using its frequency-response $H(\omega)$, or more conveniently (via a loop until convergence) using its impulse response $h(n)$, as a consequence of Parseval's theorem [33],[34], i.e.

$$\text{WNG} = \frac{1}{2\pi}\int_{-\pi}^{+\pi}|H(\omega)|^2 d\omega = \frac{1}{2\pi}\|H(\omega)\|_2^2$$

$$= \sum_{n=0}^{\infty}|h(n)|^2 = \|h(n)\|_2^2 \quad (10)$$

where $\|\cdot\|_2$ is the $\mathcal{L}_2$-norm. The gain of the maneuver error signal (MESG) is an analogous and complementary narrow-band metric. Assuming $D = 0$,

$$\text{MESG} = |H_d(\omega_{\text{man}}) - H(\omega_{\text{man}})|^2 \quad (11)$$

where $\omega_{\text{man}} = \pm\Omega T_s$ or any other frequency of interest in the passband. Now let $\sigma_{\text{tgt}}^2$ and $\sigma_{\text{man}}^2$ be the expected squared-distance errors at SS in a Cartesian space with two position coordinates $(\bar{x}, \bar{y})$ – i.e. $\sigma_{\text{sig}}^2 = \langle \varepsilon_{\bar{x}}^2 + \varepsilon_{\bar{y}}^2 \rangle$, where $\langle \cdot \rangle$ is the expectation operator and where bar accents are used to distinguish these coordinate variables from the input and output variables – for a signal produced by the target process model and a target executing a sustained constant-g turn, respectively. Furthermore, assume that white measurement-noise is added with a variance of $\sigma_{\text{sns}}^2$ in the former case. Then

$$\sigma_{\text{tgt}}^2 = 2\text{WNG} \cdot \sigma_{\text{sns}}^2 \text{ and } \sigma_{\text{man}}^2 = \text{MESG} \cdot R^2. \quad (12)$$

Note that for an SS-KF with $K = 3$ and $q = -1$: WNG above is the same as $\rho_p^2$ in (18) of [18] and $\sigma_{man}$ above is the same as $e$ in (24) of [18]. The maneuver error has components

$$\varepsilon_R = \{|H(\omega_{man})| - |H_d(\omega_{man})|\}R \text{ and}$$
$$\varepsilon_\theta = \angle H(\omega_{man}) - \angle H_d(\omega_{man}) \qquad (13)$$

where orbital parameters $\varepsilon_R$ and $\varepsilon_\theta$ are the radial and angular errors on the circular trajectory at SS. Note that the metrics in (11)-(13) all consider the desired response $H_d$, which incorporates, and compensates for, the filter lag induced by $q$.

The application of constraints at specific frequencies (via the specification of the assumed process models) may cause the magnitude response of the filter to "bulge" at intervening non-design frequencies. The resulting maxima, typically near dc, in the passband or at the passband edge, may be excited by white noise, which results in low-frequency oscillation or track "wobble". The WNG and MESG metrics are used to quantify the expected squared-distance error; however, the location $\omega_{max}$, and magnitude of the maximum, i.e. $\|H(\omega)\|_\infty^2$, are an indication of the "color" and severity of the residual errors at SS, where $\|\cdot\|_\infty$ is the $\mathcal{L}_\infty$-norm.

### III. DISCUSSION

#### A. Application

The observers and filters described in this paper may be suitable for point-target tracking and low-level image-processing functions (respectively) within a computer-vision system, for video moving-target indication, on an airborne platform [32],[35],[36]. Fixed-gain trackers are suitable in these systems, for the following reasons: High data-rates (e.g. 1080p @25Hz) and rich target/clutter environments impose heavy computational loads; The high frame-rate means that SS is reached in less than 1 second, for tracks that may last many minutes; $T_s$ is constant; Process-noise statistics are difficult to model, due to the diversity of target types and observation geometries; Process- and measurement-noise statistics are correlated, due to variable illumination, target aspect and low $T_s$. In these systems, it is also assumed that there is scope for improvement using filters with $q > 0$ (i.e. fixed-lag smoothers), as processing latencies of a few frames are likely to be imperceptible, due to the high frame-rate and other delays that exist in the system. As data association is not considered here, the tracking filters are not suitable for low-level track-processing; however, they are well suited to other high-level track-management functions, e.g. for reporting, display, classification, geographic registration, sensor handover, and sensor fusion, where the identities of inputs are known.

*B. Simulation and parameterization*

Monte Carlo (MC) simulations were conducted to investigate and illustrate the behavior of the proposed filters in this context; see Table I for some example parameterizations and see Appendix B for a complete derivation of the Filter B coefficients. The results of an MC simulation are provided in Fig. 1. In this example (Scenario 1) The target's apparent $\bar{y}$ position is displaced by 10 pixels (pix) at $n = 24$, due to a coordinate-registration update; it executes a constant radius ($R = 10$ pix) turn, with $\Omega = 2.5$ rad/s, from $n = 75$ to $n = 99$; it abruptly changes heading at $n = 125$; its apparent $\bar{y}$ position is displaced by $\pm 10$ pix on alternating frames (jitter), from $n = 160$ to $n = N_{\text{frm}} - 1$, where $N_{\text{frm}} = 190$, while it is updated by two misaligned sensors. It maintains a constant speed throughout, of 25 pix/s = 1 pix/frm @25 Hz, where $T_s = 0.04$ s. Gaussian measurement noise, with a mean of zero and a standard deviation of $\sigma_{\text{sns}} = 1$ pix is added to the target position in each frame.

The root-mean-squared distance error ($\tilde{\sigma}_D$) over all 190 frames ($N_{\text{frm}} = 190$) for 100 MC repetitions ($N_{\text{rep}} = 100$) of this scenario, for Filters A, B & C is 0.89, 1.38 & 0.82 pix, respectively ($\tilde{\sigma}_D^2 = \sum(\varepsilon_{\bar{x}}^2 + \varepsilon_{\bar{y}}^2)/N_{\text{rep}}N_{\text{frm}}$ where the $\varepsilon_{\bar{x}}$ and $\varepsilon_{\bar{y}}$ are position estimation errors using a truth track, delayed by $q$ frames). Filters B & C both have improved jitter performance relative to Filter A, which is a consequence of the interference model ($K_{\text{int}} > 0$). The degraded performance elsewhere for Filter B is restored by the maneuver model ($K_{\text{man}} > 0$) in Filter C.

For filter A (i.e. the steady-state KF), varying the process noise by multiplying $\sigma_Q$ by a factor of 8, 4, 2, 1, 0.5, 0.25 & 0.125 for a pole radius of 0.54, 0.64, 0.73, 0.8, 0.85, 0.89 & 0.92, resulted in $\tilde{\sigma}_D$ errors of 1.71, 0.82, 0.83, 0.89, 0.96, 1.21 & 1.92 (pix) respectively. On the one hand, reducing $\sigma_Q$ decreases the Nyquist gain of the KF for somewhat improved jitter tracking; however, it degrades maneuver handling, which is the dominant source of error in Scenario 1; therefore, the overall error increases. On the other hand, increasing $\sigma_Q$ improves maneuver handling; however, when taken to the extreme, the high-frequency noise-gain becomes excessive and jitter errors dominate. Filter B was altered in a similar fashion, using $p = 0.2, 0.4, 0.6, 0.7, 0.8, 0.85$ & 0.9, for $\tilde{\sigma}_D$ errors of 0.83, 0.80, 0.78, 0.83, 1.38, 2.33 & 5.01 (pix) respectively. In this case, the use of a small pole radius for a compact impulse/transient response and good maneuver handling is feasible, because the Nyquist null, introduced by the interference model, allows a wider bandwidth to be used without excessive noise amplification.

Fig. 1 also indicates that the tracks produced by a standard variable gain KF (thin red line) and the SS-KF (i.e. Filter A, thick red line) are negligible. Slight differences in the position state estimates are only visible in the first few updates. A numerical check confirms that the position element of the Kalman gain vector converges to within 0.1% of the steady-state value (of 0.36) after just 15 updates (i.e. 0.6 s) in this scenario.

*C. Response analysis*

The frequency response of the SS tracking filters is provided in Fig. 2. Note the Nyquist null for Filters B & C and the slower roll-off for Filter C. The increased high-frequency noise attenuation of these filters, relative to Filter A, is achieved at the expense of an increase in $\|H(\omega)\|_\infty^2$. Note that the SS-KF may also be configured for a Nyquist null, if $\sigma_R$ and $\sigma_Q$ are repurposed and chosen in a way that places a zero at $z = -1$. The insets expand the response (magnitude and phase error) around $f_{\text{man}}$. Note the passband distortion for Filter B and the passband flattening/broadening for Filter C. A positive magnitude (dB) and a negative phase error (e.g. B), indicate that the SS track has a larger radius and a lagging angle w.r.t the target's orbit – as quantified by $\varepsilon_R$ & $\varepsilon_\theta$ in (13).

The significance of these orbital parameters is illustrated in Fig. 3, in an additional simulation scenario, featuring only the circular maneuver with no measurement noise added (Scenario 2). This figure includes alternative configurations of Filter B, with larger and smaller pole radii. The responses of these alternative parameterizations are compared in Fig. 4. The magnitude response indicates that the radius of the SS track produced by filter B with $p = 0.9$ is larger than the true orbit for the baseline turn ($R = 10$ pix) but smaller for a tighter turn ($R = 5$ pix, see inset of Fig. 3). For these turns, the phase error response indicates that the track lags by approximately 45° and 80°, respectively. The frequency response in Fig. 4 also shows that the SS noise-attenuation improves with a larger $p$ (particularly for high frequencies) whereas the transient response shows that abrupt maneuver tracking improves with a smaller $p$ – the well-known random-error versus bias-error trade-off.

In the design of optimal infinite-impulse-response (IIR) filters for digital signal processing applications, the pole radius is typically a free parameter, subject to a stability constraint. This is reasonable because the signal parameters of interest are approximately stationary over the considered timescales (i.e. a relatively short sampling period and a relatively long application period). However, in target tracking problems this is usually not the case, thus the pole radius is fixed (i.e. placed arbitrarily) in the procedure described here, to ensure that the transient response is satisfactory – for track initiation and abrupt target manoeuvres. Constraints on the frequency response are then satisfied, for satisfactory SS behaviour – e.g. accurate kinematic state estimation, high-frequency noise attenuation and circular manoeuvre tracking – by the appropriate placement of filter zeros. In this paper, they are placed indirectly through the selection of physical process models, which maintains a connection with the kinematic states; whereas a linear combination of analytical basis functions is used in [2] & [32].

An additional simulation scenario, with all maneuver and jitter events removed (Scenario 3), was processed to assess the impact of white noise on the SS position estimate. Terminal tracking errors (at $n = 189$) for Filters A-C in Scenarios 2

& 3, were evaluated, using true positions delayed by $q$ samples, and compared with the predicted values computed using (11)-(13). Using only the final estimate in this analysis reduces the impact of the startup transient. The results are presented in Table II and in most cases the predicted values and observed values (in italics) agree to within 1%.

Up to a point, almost all aspects of a tracking filter's performance are improved by increasing $q$. These gains are of course made at the expense of the system's latency; indeed, it is reasonable to expect greater accuracy if estimates are deferred. Introducing a delay of $q$ samples results in a modulation of the response by $e^{-iq\omega}$ in the frequency domain; its effect on tracking behavior in the time domain is somewhat subtler but readily quantified and appreciated using the material presented in Section II.G.

The random-error/bias-error trade-off is an unavoidable aspect of tracker design. The proposed maneuver and interference models provide a convenient way of reaching a desirable balance that would be difficult to achieve using a standard SS-KF parameterized using Gaussian statistics. In both approaches, bias errors during sustained maneuvers are reduced by increasing filter bandwidth, which increases random errors due to measurement noise, as indicated in (10). In the proposed approach, the filter may be tuned to precisely match a deterministic circular turn. This does not necessarily degrade tracking performance for mismatched turns or result in bias errors for non-maneuvering targets. The magnitude and phase error of $H(\omega)$ at the maneuver turn frequency are constrained to be unity and zero (respectively) at the specified response frequency. The insets of Fig. 2 and the contents of Table II indicate that the satisfaction of this constraint (in Filter C) shifts the band-edge bulge (of Filter B) to a higher frequency and increases the WNG by 1 dB. At those frequencies, this distortion corresponds to: an increase in the steady-state bias errors for mismatched circular maneuvers, which are readily quantified for any circular turn of interest using (13); and an increase in the colored-noise gain. For extreme turns at high frequencies, the resulting $\|H(\omega)\|_2^2$ (i.e. the WNG) and $\|H(\omega)\|_\infty^2$ values may be unreasonably large; therefore, caution is required. Inclusion of the integrating target process model ensures that the required constraints on the frequency response at dc in (9) are satisfied, for bias-free estimation of position derivatives at steady-state, in the absence of target maneuvers, regardless of the maneuver model assumed. Fig. 2 and Table II also indicate that the high-frequency stopband attenuation and WNG reduction (of Filter B) is achieved at the expense of some low-frequency passband distortion (relative to Filter A).

TABLE I: Filter Parameters[*]

| | |
|---|---|
| **Filter A**[†]: $K = 2$, with $q = 2$, $\alpha = 0.36$ and $\beta = 0.08$ and $\boldsymbol{b} = [\alpha - \beta q, \quad \beta(1+q) - \alpha, \quad 0]$ and $\boldsymbol{a} = [1, \quad \alpha + \beta - 2, \quad 1 - \alpha]$ | |
| **Filter B**[‡]: $K = 3$, $K_{\text{tgt}} = 2, K_{\text{man}} = 0, K_{\text{int}} = 1, q = 2$ with $\boldsymbol{b} = [0.046 \quad 0.004 \quad -0.042 \quad 0]$ and $\boldsymbol{a} = [1 \quad -2.4 \quad 1.92 \quad -0.512]$ | |
| **Filter C**[‡]: $K = 5$, $K_{\text{tgt}} = 2, K_{\text{man}} = 1, K_{\text{int}} = 1, q = 2$ with $\boldsymbol{b} = [0.0899 \quad -0.1532 \quad -0.0232 \quad 0.1534 \quad -0.0666 \quad 0]$ and $\boldsymbol{a} = [1 \quad -4.0 \quad 6.4 \quad -5.12 \quad 2.048 \quad -0.3277]$ | |

[*]Independent filters designed for each spatial dimension.
[†]A 2nd-order SS-KF. This reference implementation was tuned using $\sigma_R = 1.0$ pixels (pix) and $\sigma_Q = R\Omega^2 = 62.5$ pix/s², respectively.
[‡]Designed with $\boldsymbol{a}$ extracted from $\mathcal{A}(z)$. Same pole radius as the SS-KF (i.e. $p = 0.8$), to facilitate comparison.

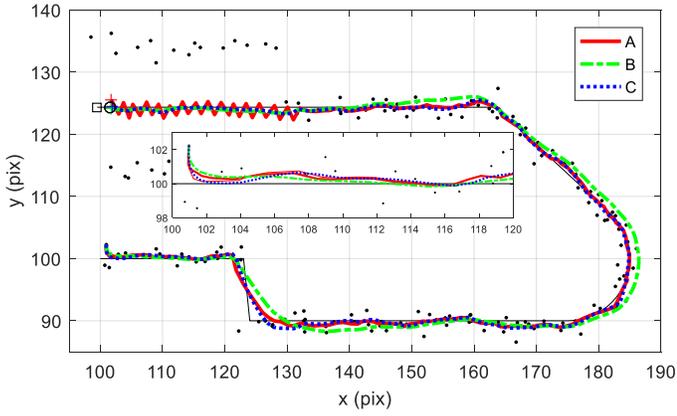

**Fig. 1**. An MC simulation instantiation showing: True target track, in pixel coordinates (black line); Final target position (black square); Final target position delayed $q$ by frames (black circle); Noisy target measurements (black dots); Track estimates (colored lines). First 20 updates shown in the inset; the variable-gain KF track is visible on this scale (thin red line).

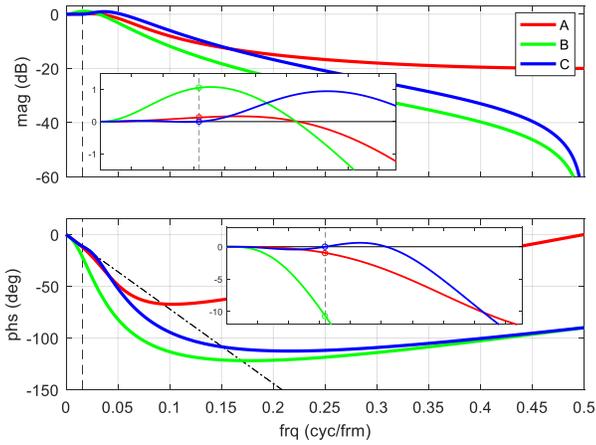

**Fig. 2**. Filter frequency response, as a function of $f = \omega/2\pi$ (cycles per frame). Maneuver frequency ($f_{\text{man}} = T_s\Omega/2\pi$) and ideal passband linear-phase response (for $q = 2$) shown.

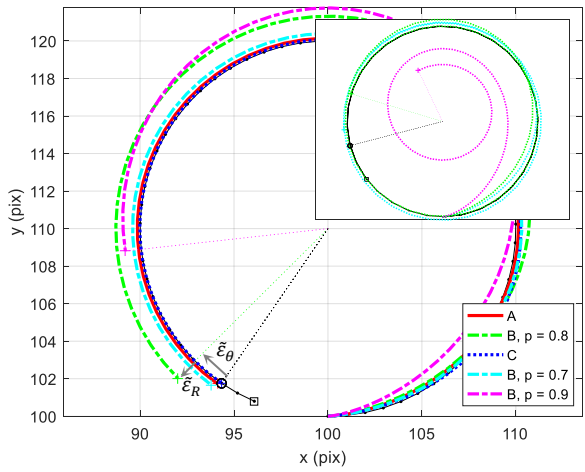

**Fig. 3**. Maneuver response, for Scenario 2 with $\Omega = 2.5$ rad/s and $R = 10$ pix (i.e. at $f_{man}$ in Fig. 2) and no measurement noise added. The $\tilde{\varepsilon}_R$ & $\tilde{\varepsilon}_\theta$ errors are shown at this time, for Filter B with $p = 0.8$. At the time of this frame ($n = 59$), Filter B with $p = 0.9$ has yet to reach SS. The SS response of this filter (up to $n = 189$) for a tighter turn ($R = 5$ pix) is shown in the inset.

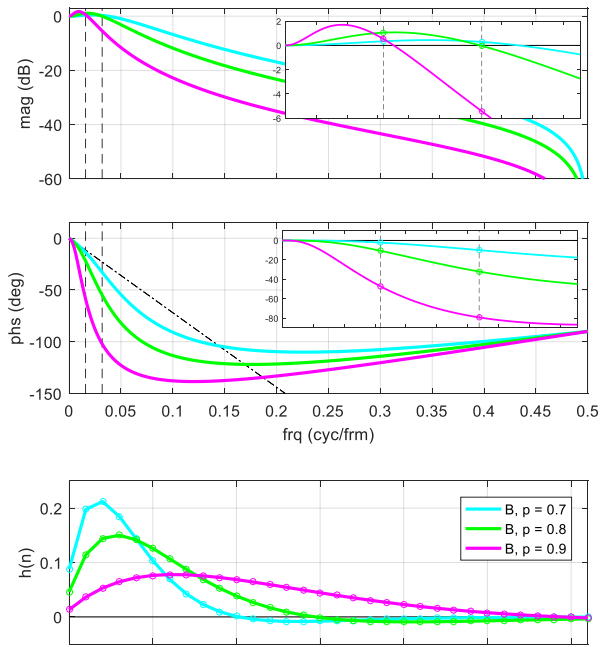

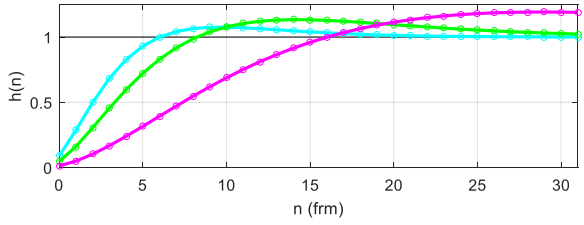

**Fig. 4**. Magnitude, phase (& phase error), unit-impulse and unit-step response (from top to bottom) as a function of pole position for Filter B with $p = 0.7, 0.8$ & $0.9$. Frequencies of baseline and tight turns shown.

TABLE II: METRIC VALIDATION

| Filter | A | B | C | B | B |
|---|---|---|---|---|---|
| $p$ | 0.8 | 0.8 | 0.8 | 0.7 | 0.9 |
| MESG (dB) | -33.06 | -12.55 | -187.2 | -25.07 | -1.615 |
| MESG | 4.9e-4 | 5.6e-2 | 1.9e-19 | 3.1e-3 | 0.689 |
| $\sigma_{man}$ (pix) | 0.222 | 2.358 | 4.4e-9 | 0.558 | 8.303 |
| $\tilde{\sigma}_D$ (pix) [a] | 0.221 | 2.345 | 1.6e-3 | 0.555 | 8.284 |
| $\varepsilon_R$ (pix) | 0.148 | 1.288 | -2.3e-9 | 0.400 | 0.624 |
| $\tilde{\varepsilon}_R$ (pix) [a] | 0.147 | 1.287 | -8.1e-4 | 0.398 | 0.643 |
| $\varepsilon_\theta$ (°) | -0.945 | -10.66 | -2.1e-8 | -2.186 | -47.36 |
| $\tilde{\varepsilon}_\theta$ (°) [a] | -0.936 | -10.59 | -7.8e-3 | -2.168 | -47.19 |
| WNG (dB) | -8.081 | -9.043 | -7.254 | -7.823 | -11.61 |
| WNG | 0.156 | 0.125 | 0.188 | 0.165 | 0.069 |
| $\sigma_{tgt}$ (pix) | 0.558 | 0.499 | 0.614 | 0.575 | 0.372 |
| $\tilde{\sigma}_D$ (pix) [b] | 0.559 | 0.499 | 0.612 | 0.574 | 0.371 |

[a] Terminal error in Scenario 2.
[b] Terminal error in Scenario 3, averaged over 5000 random instantiations.

## IV. CONCLUSION

Tracker development is an exercise in compromise. Tracker design by pole placement is useful because it emphasizes convergence behavior, which allows the balance between the transient response and the steady-state response of a tracker to be directly determined. Similar responses may also be achieved using a SS Kalman filter; indeed, the response is optimal for a given combination of process-noise and measurement-noise parameters; however, the assumed statistical distributions are rarely a reasonable approximation of operational reality (all of the time); therefore, the designer of a tracking system is forced to empirically tune the tracker for the desired transient and steady-state response by manually adjusting the statistical parameters until the tracker requirements are met and the system operators are satisfied. Fixed-gain trackers designed by pole placement obviate the need for noise statistics and focus on system requirements/expectations directly.

The relationship between a tracking filter's frequency response and its tracking behavior at steady-state has been summarized and discussed; namely, the link between: a) The complex derivatives at the dc limit, the kinematic state derivatives, and the latency of the filter; b) The $\mathcal{L}_2$-norm in the frequency domain and the measurement-noise attenuation; c) The complex response at a given frequency and the (radial, angular and distance) errors during a constant-g maneuver on a circle with a given radius. While similar metrics have been used elsewhere in the literature to analyze $\alpha - \beta(-\gamma)$ filters, a design method that specifically addresses these requirements is described in this paper.

The simulation results and analysis indicate that the proposed augmented-state observer provides a simple and intuitive way of balancing the various (transient and SS) performance trade-offs associated with the design of fixed-gain tracking filters, that would otherwise be difficult to achieve using a standard SS-KF with a filter order equal to the Newtonian state dimension.

## ACKNOWLEDGEMENTS

I would like to thank N. J. Redding for taking an interest in this work and J. L. Williams for introducing me to shaping filters.

## APPENDICES

### A. Design by pole placement

The poles of $\mathcal{H}(z)$ are equal to the roots of the denominator polynomial

$$\mathcal{A}(z) = |z\mathbf{I} - \mathbf{G}_{\text{obs}}| \quad (A1)$$

which yields the so-called characteristic equation of the LSS system

$$z^K + \sum_{k=1}^{K} \mathbf{a}(k) z^{K-k} = 0 \quad (A2)$$

that may be solved by finding the eigenvalues of $\mathbf{G}_{\text{obs}}$.

Given the complexity of (A2) for the ($\mathcal{Z}$-transformed) LSS observer system defined in (5), which is a polynomial containing the scrambled elements of $\mathcal{K}$, it is not clear how the elements of $\mathcal{K}$ should be chosen in this coordinate system to place the observer poles at the desired locations. To address this problem, a new coordinate system is used. The new *canonical* coordinates are a linear combination of the old *kinematic* coordinates, i.e.

$$\mathbf{w}_{\text{obs}}^{\text{pcf}} = \mathbb{T}_{\text{prc}}^{\text{pcf}\leftarrow\text{kin}} \mathbf{w}_{\text{prc}}^{\text{kin}}; \text{ thus, in (5):}$$

$G_{\text{obs}}^{\text{kin}} = G_{\text{prc}}^{\text{kin}} - \mathcal{K}^{\text{kin}} C_{\text{prd}}^{\text{kin}}$ and $H_{\text{obs}}^{\text{kin}} = \mathcal{K}^{\text{kin}}$ become

$G_{\text{obs}}^{\text{pcf}} = G_{\text{prc}}^{\text{pcf}} - \mathcal{K}^{\text{pcf}} C_{\text{prd}}^{\text{pcf}}$ and $H_{\text{obs}}^{\text{pcf}} = \mathcal{K}^{\text{pcf}}$ where     (A3)

$\mathcal{K}^{\text{kin}}$ is the gain vector in the kinematic coordinate system and

$\mathcal{K}^{\text{pcf}}$ is the gain vector in the canonical coordinate system.

(Superscripts are used to identify the coordinate system.)

The linear transformation $\mathbb{T}_{\text{prc}}^{\text{pcf} \leftarrow \text{kin}}$, preserves the eigenvalues (i.e. the poles) of $G_{\text{prc}}^{\text{kin}}$ and it is specifically formulated so that it reduces the $\langle C_{\text{prd}}^{\text{kin}}, G_{\text{prc}}^{\text{kin}} \rangle$ observable pair to what is referred to here as process canonical form (PCF), which is (by definition):

$$G_{\text{prc}}^{\text{pcf}} = \begin{bmatrix} 0 & 0 & \cdots & 0 & 0 & g_{\text{prc}}(0) \\ 1 & 0 & \cdots & 0 & 0 & g_{\text{prc}}(1) \\ 0 & 1 & \cdots & 0 & 0 & g_{\text{prc}}(2) \\ \vdots & \vdots & \ddots & \vdots & \vdots & \vdots \\ 0 & 0 & \cdots & 1 & 0 & g_{\text{prc}}(K-2) \\ 0 & 0 & \cdots & 0 & 1 & g_{\text{prc}}(K-1) \end{bmatrix}_{K \times K} = \begin{bmatrix} \mathbf{0}_{1 \times (K-1)} & g_{\text{prc}} \\ I_{(K-1)} & \end{bmatrix}_{K \times K}$$

with

$$C_{\text{prd}}^{\text{pcf}} = [0 \quad 0 \quad 0 \quad \cdots \quad 0 \quad 1]_{1 \times K} = [\mathbf{0}_{1 \times (K-1)} \quad 1]_{1 \times K}. \qquad (A4)$$

This form is ideal because the eigenvalues of $G_{\text{prc}}^{\text{pcf}}$ are readily extracted from its $K$th column; this form is also ideal because only the $K$th element of $C_{\text{prd}}^{\text{pcf}}$ is non-zero. The latter feature ensures that $\mathcal{K}^{\text{pcf}}$ only appears in the $K$th column of $G_{\text{obs}}^{\text{pcf}}$. This means that $\mathcal{K}^{\text{pcf}}$ may now be used to determine the response of the observer. The transient response is important because it determines how the observer responds to abrupt changes in the signal states, e.g. brief maneuvers (in target-tracking systems) or edges (in computer-vision systems). The steady-state (i.e. frequency) response is important because it determines how the observer responds to sustained constant-g turns and white noise at steady state. Using $p \to 0$ improves the transient response; whereas $p \to 1$ improves the steady-state response; for example, using $p = 0.8$ was found to be a reasonable compromise for Scenario 1 in the Discussion.

In PCF, the state-transition matrix of the observer in (A3) becomes

$$G_{\text{obs}}^{\text{pcf}} = \begin{bmatrix} \mathbf{0}_{1 \times (K-1)} & g_{\text{prc}} - \mathcal{K}^{\text{pcf}} \\ I_{(K-1)} & \end{bmatrix}_{K \times K}. \text{ Now let}$$

$g_{\text{obs}}(k) = g_{\text{prc}}(k) - \mathcal{K}^{\text{pcf}}(k)$ or $\boldsymbol{g}_{\text{obs}} = \boldsymbol{g}_{\text{prc}} - \boldsymbol{\mathcal{K}}^{\text{pcf}}$ so that

$$\boldsymbol{G}_{\text{obs}}^{\text{pcf}} = \begin{bmatrix} \boldsymbol{0}_{1\times(K-1)} & \\ \boldsymbol{I}_{(K-1)} & \boldsymbol{g}_{\text{obs}} \end{bmatrix}_{K\times K} \quad (A5)$$

In PCF, the effect that $\boldsymbol{\mathcal{K}}^{\text{pcf}}$ has on the poles of $\boldsymbol{G}_{\text{obs}}^{\text{pcf}}$ is now evident. For a matrix with the canonical structure of $\boldsymbol{G}_{\text{obs}}^{\text{pcf}}$ (or $\boldsymbol{G}_{\text{prc}}^{\text{pcf}}$), the eigenvalues are equal to the roots of a polynomial formed from its $K$th column as follows:

$$z^K - \sum_{k=0}^{K-1} g_{\text{obs}}(k) z^k = \prod_{k=0}^{K-1}(z - \lambda_k). \quad (A6)$$

Let the discrete-time transfer function relating $x$ to $y$, i.e. $\mathcal{H}(z)$ be the ratio of two polynomials $\mathcal{B}_{\text{obs}}(z)$ and $\mathcal{A}_{\text{obs}}(z)$ that determine the zeros and poles of the observer, respectively, i.e. $H(z) = \mathcal{B}_{\text{obs}}(z)/\mathcal{A}_{\text{obs}}(z)$. As the eigenvalues of $\boldsymbol{G}_{\text{obs}}^{\text{pcf}}$ are equal to the poles of the observer, we have

$\mathcal{A}_{\text{obs}}(z) = z^K - \sum_{k=0}^{K-1} g_{\text{obs}}(k) z^k = z^K + \sum_{k=1}^{K} a_{\text{obs}}(k) z^{K-k}$ thus

$a_{\text{obs}}(k) = -g_{\text{obs}}(K - k)$, for $1 \leq k \leq K$ or

$g_{\text{obs}}(k) = -a_{\text{obs}}(K - k)$, for $0 \leq k < K$. $\quad (A7)$

i.e. opposite sign with indexing reversed and offset by one place.

In PCF it is apparent that: The elements of $\boldsymbol{\mathcal{K}}^{\text{pcf}}$ are simply chosen to yield an observer polynomial $\mathcal{A}_{\text{obs}}(z)$, that has roots at the desired pole locations $\lambda_k$, $(= p)$; given the process polynomial $\mathcal{A}_{\text{prc}}(z)$, with roots at the pole locations $\rho_k$, for $0 \leq k < K$, as specified in the process model in (4). Thus the gain vector is determined using

$\boldsymbol{\mathcal{K}}^{\text{pcf}} = \boldsymbol{g}_{\text{prc}} - \boldsymbol{g}_{\text{obs}}$ where $\quad (A8)$

$\boldsymbol{g}_{\text{prc}}$ is extracted from $\boldsymbol{G}_{\text{prc}}^{\text{pcf}}$ as shown in (A4) and

$\boldsymbol{g}_{\text{obs}}$ is extracted from $\boldsymbol{G}_{\text{obs}}^{\text{pcf}}$ as shown in (A5).

The elements of $\boldsymbol{g}_{\text{obs}}$ are obtained from the poles of the observer, which are arbitrarily set for the desired convergence properties; whereas the elements of $\boldsymbol{g}_{\text{prc}}$ are obtained from the poles of the process model, which have been set to represent the dynamics of the physical processes that generate the sensor measurement sequence.

With $\mathcal{K}^{\text{pcf}}$ determined in canonical coordinates, it is transformed back into kinematic coordinates, using

$$\mathcal{K}^{\text{kin}} = \mathbb{T}_{\text{prc}}^{\text{kin}\leftarrow\text{pcf}} \mathcal{K}^{\text{pcf}}. \tag{A9}$$

It has so far been assumed that the $\mathbb{T}_{\text{prc}}^{\text{pcf}\leftarrow\text{kin}}$ transform and its inverse $\mathbb{T}_{\text{prc}}^{\text{kin}\leftarrow\text{pcf}} = \left\{\mathbb{T}_{\text{prc}}^{\text{pcf}\leftarrow\text{kin}}\right\}^{-1}$ are known. It is clearly possible to compute $\mathcal{K}^{\text{pcf}}$ without this knowledge; however, it is not possible to realize the observer, because the $C_{\text{obs}}^{\text{pcf}}$ operator for this coordinate system is unknown. Fortunately, the required transforms are readily found using

$$\mathbb{T}_{\text{prc}}^{\text{kin}\leftarrow\text{pcf}} = \left\{\mathcal{O}_{\text{prc}}^{\text{kin}}\right\}^{-1} \mathcal{O}_{\text{prc}}^{\text{pcf}} \tag{A10}$$

where $\mathcal{O}$ is the $K \times K$ observability matrix for the $\langle C, G \rangle$ pair, with the $k$th row equal to $CG^k$ (for $0 \leq k < K$). The observability matrices for the kinematic and PCF systems are constructed using the $\langle C_{\text{prd}}^{\text{kin}}, G_{\text{prc}}^{\text{kin}} \rangle$ and $\langle C_{\text{prd}}^{\text{pcf}}, G_{\text{prc}}^{\text{pcf}} \rangle$ definitions provided above in (5) and (A4), respectively. The transform only exists if $\langle C_{\text{prd}}^{\text{kin}}, G_{\text{prc}}^{\text{kin}} \rangle$ is observable, i.e. if $\text{rank}\{\mathcal{O}_{\text{prc}}^{\text{kin}}\} = K$.

The kin ← pcf coordinate transform applied using $\mathbb{T}_{\text{prc}}^{\text{kin}\leftarrow\text{pcf}}$ was used to expose the coefficients of the $\mathcal{A}_{\text{prc}}(z)$ polynomial in the $G_{\text{prc}}$ matrix and to focus the action of the $C_{\text{prd}}$ operator so that the gain vector $\mathcal{K}$ only acts on the desired column of $G_{\text{prc}}$. A similar transformation is applied below to expose the coefficients of the $\mathcal{B}_{\text{obs}}(z)$ polynomial in the $H_{\text{obs}}$ operator.

The kinematic observer equations in (5) are not in a canonical form after the gain vector is introduced to form $G_{\text{obs}}^{\text{kin}}$ and when the $q$-parameterized $C_{\text{obs}}^{\text{kin}}$ operator is used; therefore, a new transform $\mathbb{T}_{\text{obs}}^{\text{kin}\leftarrow\text{ocf}}$ is required. It is computed using a procedure that is analogous to the one that was used previously to compute $\mathbb{T}_{\text{prc}}^{\text{kin}\leftarrow\text{pcf}}$; with observability matrices for the observable $\langle G_{\text{obs}}^{\text{kin}}, H_{\text{obs}}^{\text{kin}} \rangle$ pair now used instead of observability matrices for the observable $\langle C_{\text{prc}}^{\text{kin}}, G_{\text{prc}}^{\text{kin}} \rangle$ pair.

The $\mathbb{T}_{\text{obs}}^{\text{ocf}\leftarrow\text{kin}}$ operator changes the kinematic coordinate system in a way that transforms the observable $\langle C_{\text{obs}}^{\text{kin}}, G_{\text{obs}}^{\text{kin}} \rangle$ pair into a canonical $\langle C_{\text{obs}}^{\text{ocf}}, G_{\text{obs}}^{\text{ocf}} \rangle$ pair with an observer canonical form (OCF); the $\mathbb{T}_{\text{obs}}^{\text{kin}\leftarrow\text{ocf}}$ operator reverses the transformation, i.e.

$$w_{\text{obs}}^{\text{ocf}} = \mathbb{T}_{\text{obs}}^{\text{ocf}\leftarrow\text{kin}} w_{\text{obs}}^{\text{kin}} \text{ and } w_{\text{obs}}^{\text{kin}} = \mathbb{T}_{\text{obs}}^{\text{kin}\leftarrow\text{ocf}} w_{\text{obs}}^{\text{ocf}} \text{ such that} \tag{6.3.1b}$$

$$\langle C_{\text{obs}}^{\text{kin}}, G_{\text{obs}}^{\text{kin}} \rangle \underset{\mathbb{T}_{\text{obs}}^{\text{ocf}\leftarrow\text{kin}}}{\overset{\mathbb{T}_{\text{obs}}^{\text{kin}\leftarrow\text{ocf}}}{\rightleftarrows}} \langle C_{\text{obs}}^{\text{ocf}}, G_{\text{obs}}^{\text{ocf}} \rangle \text{ with}$$

$$\mathbb{T}_{\text{obs}}^{\text{kin}\leftarrow\text{ocf}} = \{\mathcal{O}_{\text{obs}}^{\text{kin}}\}^{-1}\mathcal{O}_{\text{obs}}^{\text{ocf}} \text{ and } \mathbb{T}_{\text{obs}}^{\text{ocf}\leftarrow\text{kin}} = \{\mathbb{T}_{\text{obs}}^{\text{kin}\leftarrow\text{ocf}}\}^{-1}. \quad \text{(A11)}$$

In this canonical coordinate system, we have the following:

$$w_{\text{obs}}^{\text{ocf}}(n) = G_{\text{obs}}^{\text{ocf}} w_{\text{obs}}^{\text{ocf}}(n-1) + H_{\text{obs}}^{\text{ocf}} x(n)$$

$$y(n) = C_{\text{obs}}^{\text{ocf}} w_{\text{obs}}^{\text{ocf}}(n) \text{ where}$$

$$G_{\text{obs}}^{\text{ocf}} = \begin{bmatrix} \mathbf{0}_{1\times(K-1)} & \\ I_{(K-1)} & g_{\text{obs}} \end{bmatrix}_{K\times K}, H_{\text{obs}}^{\text{ocf}} = \mathbb{T}_{\text{obs}}^{\text{ocf}\leftarrow\text{kin}} H_{\text{obs}}^{\text{kin}}$$

$$C_{\text{obs}}^{\text{ocf}} = [\mathbf{0}_{1\times(K-1)} \quad 1]_{1\times K} \text{ (by definition)}$$

$$w_{\text{obs}}^{\text{ocf}}(0) = \mathbb{T}_{\text{obs}}^{\text{ocf}\leftarrow\text{kin}} w_{\text{obs}}^{\text{kin}}(0) \text{ for state initialization and}$$

$$w_{\text{obs}}^{\text{kin}}(n) = \mathbb{T}_{\text{obs}}^{\text{kin}\leftarrow\text{ocf}} w_{\text{obs}}^{\text{ocf}}(n) \text{ for state extraction.} \quad \text{(A12)}$$

The correctness of the transform may be confirmed using:

$$G_{\text{obs}}^{\text{ocf}} = \mathbb{T}_{\text{obs}}^{\text{ocf}\leftarrow\text{kin}} G_{\text{obs}}^{\text{kin}} \mathbb{T}_{\text{obs}}^{\text{kin}\leftarrow\text{ocf}} \text{ and } C_{\text{obs}}^{\text{ocf}} = C_{\text{obs}}^{\text{kin}} \mathbb{T}_{\text{obs}}^{\text{kin}\leftarrow\text{ocf}}.$$

For a strictly proper $\mathcal{H}(z)$ where the order of $\mathcal{B}_{\text{obs}}(z)$ is less than the order of $\mathcal{A}_{\text{obs}}(z)$, the coefficients of the $\mathcal{B}_{\text{obs}}(z)$ and $\mathcal{A}_{\text{obs}}(z)$ polynomials, which define the zeros and poles of the filter respectively, are extracted from the LSS representation of the observer in OCF, as follows:

$$G_{\text{obs}}^{\text{ocf}} = \begin{bmatrix} 0 & 0 & \cdots & 0 & 0 & -a_{\text{obs}}(K-0) \\ 1 & 0 & \cdots & 0 & 0 & -a_{\text{obs}}(K-1) \\ 0 & 1 & \cdots & 0 & 0 & -a_{\text{obs}}(K-2) \\ \vdots & \vdots & \ddots & \vdots & \vdots & \vdots \\ 0 & 0 & \cdots & 1 & 0 & -a_{\text{obs}}(2) \\ 0 & 0 & \cdots & 0 & 1 & -a_{\text{obs}}(1) \end{bmatrix}_{K\times K} H_{\text{obs}}^{\text{ocf}} = \begin{bmatrix} b_{\text{obs}}(K-1) \\ b_{\text{obs}}(K-2) \\ b_{\text{obs}}(K-3) \\ \vdots \\ b_{\text{obs}}(1) \\ b_{\text{obs}}(0) \end{bmatrix}_{K\times 1}$$

$$C_{\text{obs}}^{\text{ocf}} = [0 \quad 0 \quad 0 \quad \cdots \quad 0 \quad 1]_{1\times K}. \quad \text{(A13)}$$

As exploited in (A5), the $a_{\text{obs}}$ coefficients of $\mathcal{A}_{\text{obs}}(z)$ are arranged along the last column of $G_{\text{obs}}^{\text{ocf}}$. Equation (A13) shows that the $b_{\text{obs}}$ coefficients of $\mathcal{B}_{\text{obs}}(z)$ are similarly arranged in $H_{\text{obs}}^{\text{ocf}}$. Note that the $a_{\text{obs}}(0)$ coefficient and the $b_{\text{obs}}(K)$ coefficient do not appear above, because they are required to be unity and zero, respectively, for the strictly proper systems considered here.

*B. Worked example*

A worked example, for an observer (i.e. Filter B) with $K_{\text{tgt}} = 2, K_{\text{man}} = 0, K_{\text{int}} = 1, p = 0.8, q = 2$ (samples) and $T_s = 0.04$ (seconds), is presented below. For the discrete-time model of the process, as specified in (4):

$$G_{\text{tgt}}^{\text{kin}} = \begin{bmatrix} 1 & 0.04 \\ 0 & 1 \end{bmatrix}, C_{\text{tgt}}^{\text{kin}} = \begin{bmatrix} 1 & 0 \end{bmatrix} \text{ and}$$

$$G_{\text{int}}^{\text{kin}} = -1, C_{\text{int}}^{\text{kin}} = 1 \text{ thus}$$

$$G_{\text{prc}}^{\text{kin}} = \begin{bmatrix} 1 & 0.04 & 0 \\ 0 & 1 & 0 \\ 0 & 0 & -1 \end{bmatrix}, C_{\text{prc}}^{\text{kin}} = \begin{bmatrix} 1 & 0 & 1 \end{bmatrix} \text{ with}$$

$$C_{\text{prd}}^{\text{kin}} = \begin{bmatrix} 1 & 0.04 & -1 \end{bmatrix} \text{ and } C_{\text{obs}}^{\text{kin}} = \begin{bmatrix} 1 & -0.08 & 0 \end{bmatrix}$$

For our process with system poles $\rho_0 = 1, \rho_1 = 1$ & $\rho_2 = -1$, we know that

$$\mathcal{A}_{\text{prc}}(z) = z^3 - z^2 - z + 1 \text{ thus}$$

$$\boldsymbol{a}_{\text{prc}} = \begin{bmatrix} 1 & -1 & -1 & 1 \end{bmatrix} \text{ and } \boldsymbol{g}_{\text{prc}} = \begin{bmatrix} -1 \\ 1 \\ 1 \end{bmatrix} \text{ which yields}$$

$$G_{\text{prc}}^{\text{pcf}} = \begin{bmatrix} 0 & 0 & -1 \\ 1 & 0 & 1 \\ 0 & 1 & 1 \end{bmatrix} \text{ and } C_{\text{prd}}^{\text{pcf}} = \begin{bmatrix} 0 & 0 & 1 \end{bmatrix}$$

in PCF coordinates for the $\langle C_{\text{prd}}^{\text{kin}}, G_{\text{prc}}^{\text{kin}} \rangle$ pair.

For our observer with $\lambda_k = p = 0.8$ for all $k$, we know that

$$\mathcal{A}_{\text{obs}}(z) = z^3 - 2.400 z^2 + 1.920 z - 0.512 \text{ thus}$$

$$\boldsymbol{a}_{\text{obs}} = \begin{bmatrix} 1.000 & -2.400 & 1.920 & -0.512 \end{bmatrix} \text{ and}$$

$$\boldsymbol{g}_{\text{obs}} = \begin{bmatrix} 0.512 \\ -1.920 \\ 2.400 \end{bmatrix} \text{ which yields}$$

$$G_{\text{obs}}^{\text{pcf}} = \begin{bmatrix} 0 & 0 & 0.512 \\ 1 & 0 & -1.920 \\ 0 & 1 & 2.400 \end{bmatrix}$$

in PCF coordinates for the $\langle C_{\text{prd}}^{\text{kin}}, G_{\text{prc}}^{\text{kin}} \rangle$ pair. The gain vector in PCF coordinates is

$$\mathcal{K}^{\text{pcf}} = \begin{bmatrix} -1.512 \\ 2.920 \\ -1.400 \end{bmatrix}.$$

We now need to find the transform that converts this gain vector back into kinematic coordinates. The observability matrices in both coordinate systems are

$$\mathcal{O}_{\text{prc}}^{\text{kin}} = \begin{bmatrix} 1 & 0.04 & -1 \\ 1 & 0.08 & 1 \\ 1 & 0.12 & -1 \end{bmatrix} \text{ and } \mathcal{O}_{\text{prc}}^{\text{pcf}} = \begin{bmatrix} 0 & 0 & 1 \\ 0 & 1 & 1 \\ 1 & 1 & 2 \end{bmatrix}.$$

The required transform is then found using

$$\mathbb{T}_{\text{prc}}^{\text{kin}\leftarrow\text{pcf}} = \begin{bmatrix} -0.75 & -0.25 & 0.25 \\ 12.50 & 12.50 & 12.50 \\ -0.25 & 0.25 & -0.25 \end{bmatrix} \text{ and}$$

$$\mathbb{T}_{\text{prc}}^{\text{pcf}\leftarrow\text{kin}} = \begin{bmatrix} -1 & 0.00 & -1 \\ 0 & 0.04 & 2 \\ 1 & 0.04 & -1 \end{bmatrix} \text{ thus } \mathcal{K}^{\text{kin}} = \begin{bmatrix} 0.054 \\ 0.100 \\ 1.458 \end{bmatrix}$$

$$\boldsymbol{G}_{\text{obs}}^{\text{kin}} = \begin{bmatrix} 0.9460 & 0.0378 & 0.0540 \\ -0.1000 & 0.9960 & 0.1000 \\ -1.4580 & -0.0583 & 0.4580 \end{bmatrix} \text{ and } \boldsymbol{H}_{\text{obs}}^{\text{kin}} = \mathcal{K}^{\text{kin}}.$$

We now have everything we need to realize our observer using the triplet $\langle \boldsymbol{C}_{\text{obs}}^{\text{kin}}, \boldsymbol{G}_{\text{obs}}^{\text{kin}}, \boldsymbol{H}_{\text{obs}}^{\text{kin}} \rangle$. However, to extract the $\boldsymbol{b}_{\text{obs}}$ coefficients for analysis, a conversion to OCF is required, using

$$\mathcal{O}_{\text{obs}}^{\text{kin}} = \begin{bmatrix} 1 & -0.0800 & 0 \\ 0.9540 & -0.0418 & 0.0460 \\ 0.8396 & -0.0083 & 0.0684 \end{bmatrix} \text{ and}$$

$$\mathcal{O}_{\text{obs}}^{\text{ocf}} = \begin{bmatrix} 0 & 0 & 1 \\ 0 & 1 & 2.40 \\ 1 & 2.40 & 3.84 \end{bmatrix} \text{ which yields}$$

$$\mathbb{T}_{\text{obs}}^{\text{kin}\leftarrow\text{ocf}} = \begin{bmatrix} 10.4688 & 9.5585 & 9.9012 \\ 130.8603 & 119.4811 & 111.2654 \\ -98.0883 & -67.8198 & -51.9659 \end{bmatrix} \text{ and}$$

$$\mathbb{T}_{\text{obs}}^{\text{ocf}\leftarrow\text{kin}} = \begin{bmatrix} 0.470 & -0.0614 & -0.042 \\ -1.446 & 0.1502 & 0.046 \\ 1.000 & -0.0800 & 0.000 \end{bmatrix}. \text{ Thus}$$

$$\boldsymbol{H}_{\text{obs}}^{\text{ocf}} = \begin{bmatrix} -0.042 \\ 0.004 \\ 0.046 \end{bmatrix} \text{ and } \boldsymbol{b}_{\text{obs}} = [0.046 \quad 0.004 \quad -0.042 \quad 0].$$

The observer may also be realized in OCF using the triplet $\langle \boldsymbol{C}_{\text{obs}}^{\text{ocf}}, \boldsymbol{G}_{\text{obs}}^{\text{ocf}}, \boldsymbol{H}_{\text{obs}}^{\text{ocf}} \rangle$, noting that $\boldsymbol{G}_{\text{obs}}^{\text{ocf}} = \boldsymbol{G}_{\text{obs}}^{\text{pcf}}$ and $\boldsymbol{C}_{\text{obs}}^{\text{ocf}}$ is defined in (13). Reducing the state observer to a set of $\boldsymbol{b}_{\text{obs}}$ and $\boldsymbol{a}_{\text{obs}}$ coefficients allows it to be realized using optimized code/hardware via a standard interface. The initial (non-kinematic) filter state is set by converting the initial kinematic state, using an appropriate transformation matrix, e.g. using (A11) & (A12); or by assuming that the initial position input has been applied forever. In the latter case, the internal state-vector elements are set equal to the steady-state values of their respective responses to a step input with a magnitude equal to the initial input [38]. In either case, the internal structure of the filter realization must be known, as it may not be the same as the canonical form used here.

C. *List of symbols (in approximate order of appearance)*

| | |
|---|---|
| $K$ | Dimension of state vector |
| $\boldsymbol{Q}$ | Process noise matrix |
| $\boldsymbol{R}$ | Measurement noise matrix |
| $\alpha, \beta, \gamma$ | Fixed gains for position, velocity and acceleration states |
| $\lambda_{\text{SS-KF}}$ | Tracking index |
| $T_s$ | Sampling period |
| $\sigma_Q$ | Standard deviation of process noise (assumed) |
| $\sigma_R$ | Standard deviation of measurement noise (assumed) |
| $z$ | Coordinate in the complex $z$-plane |
| $\omega$ | Angular frequency |
| $f$ | Relative frequency $f = \omega/2\pi$ |
| $t$ | Continuous time variable |
| $y_{[\cdot]}$ | System output variable of the subscripted system |
| $x_{[\cdot]}$ | System input variable of the subscripted system |
| $\mathbf{I}$ | Identity matrix |
| $\mathbf{w}_{[\cdot]}$ | State-vector of the subscripted system |
| $\mathbf{A}_{[\cdot]}$ | Continuous-time state matrix (square) for the subscripted system |
| $\mathbf{C}_{[\cdot]}$ | Output matrix (a row vector) for the subscripted system |
| $[\cdot]_{\text{tgt}}$ | Pertaining to target process model system |
| $[\cdot]_{\text{man}}$ | Pertaining to manoeuver process model system |
| $[\cdot]_{\text{int}}$ | Pertaining to interference process model system |
| $[\cdot]_{\text{prc}}$ | Pertaining to augmented process model system |
| $[\cdot]_{\text{obs}}$ | Pertaining to augmented state observer system |
| $[\cdot]_{\text{prd}}$ | Pertaining to a predicted state |
| $[\cdot]_{\text{sig}}$ | Pertaining to signal model system (manoeuver or target) |
| $\mathcal{L}\{\cdot\}$ | Laplace transform operator |
| $s$ | Coordinate in the complex $s$-plane |
| $k$ | State-vector index ($k = 0 \ldots K - 1$) |
| $M \times N$ | Generic matrix dimensions |
| $\mathbf{G}_{[\cdot]}$ | Discrete-time state-transition matrix (square) for the subscripted system |
| $\mathbf{H}_{[\cdot]}$ | Input matrix (column vector) for the subscripted system |
| $\mathcal{G}$ | Elements of $\mathbf{G}_{\text{tgt}}$, for an integrating system |
| $\boldsymbol{\Phi}_{[\cdot]}(s)$ | Laplace transform of the fundamental matrix for the subscripted system |
| $\Omega$ | Angular velocity of maneuver orbit |
| $R$ | Radius of maneuver orbit |
| $i$ | Imaginary unit $= \sqrt{-1}$ |
| $n$ | Discrete time index, $t = nT_s$ |
| $\hat{\mathbf{w}}_{[\cdot]}$ | Observer state-vector, i.e. estimate of process state |
| $\hat{x}$ | Estimate of observer input (process output) |
| $[\cdot]^T$ | Transpose operator |
| $\mathcal{K}^{[\cdot]}$ | Observer gain (column vector), in the coordinates of the superscript |
| $q$ | Delay parameter, for a time shift of $\Delta t = -qT_s$ |
| $\rho_k$ | Position of the $k$th process pole in the complex $z$-plane |
| $\lambda_k$ | Position of the $k$th observer pole in the complex $z$-plane |
| $p$ | Position of repeated observer poles on the real axis in the complex $z$-plane |
| $D$ | Derivative order of SISO observer output |
| $\mathcal{H}(z)$ | Discrete-time transfer function of SISO observer system, linking the input $x$ to the output $y$ |
| $\mathcal{B}(z)$ | Numerator polynomial of $\mathcal{H}(z)$ |
| $\mathcal{A}(z)$ | Denominator polynomial of $\mathcal{H}(z)$ |
| $\boldsymbol{b}_{[\cdot]}$ (or $b_{[\cdot]}$) | Coefficient vector (or scalar) of the $\mathcal{B}(z)$ polynomial and for the delayed input sequence in the canonical realization of $\mathcal{H}(z)$ |
| $\boldsymbol{a}_{[\cdot]}$ (or $a_{[\cdot]}$) | Coefficient vector (or scalar) of the $\mathcal{A}(z)$ polynomial and for the delayed output sequence in the canonical realization of $\mathcal{H}(z)$ |
| $\boldsymbol{\mathcal{D}}_{[\cdot]}$ | State derivative operator, $\boldsymbol{\mathcal{D}}_{\text{sig}} = \boldsymbol{A}_{\text{sig}}^D$ |
| $H(\omega)$ | Frequency response of the SISO observer $\mathcal{H}(z)$ |
| $\omega_c$ | A critical frequency, $\omega_c = 0$, $\pm\Omega T_s$ or $\pi$ |
| $L_{\text{dc}}$ | Number of (dc) derivative constraints at $\omega = 0$ |
| $L_{\text{wb}}$ | Number of (wide-band) derivative constraints at $\omega = \pm\Omega T_s$ |
| $L_{\text{pi}}$ | Number of (Nyquist) derivative constraints at $\omega = \pi$ |
| $L$ | Total number of derivative constraints |
| $\rho_l(\omega_c)$ | The required value of the $l$th frequency derivative of $H(\omega)$, evaluated at $\omega_c$ |
| $\|\cdot\|_2$ | $\mathcal{L}_2$-norm |
| $\|\cdot\|_\infty$ | $\mathcal{L}_\infty$-norm |
| $h(n)$ | Impulse response of the SISO observer $\mathcal{H}(z)$ |
| $\omega_{\max}$ | The frequency at which $|H(\omega)|$ is maximized |
| $\omega_{\text{man}}$ | Maneuver test frequency, in the passband |
| $f_{\text{man}}$ | Normalized maneuver test frequency |
| $H_d(\omega)$ | Desired frequency response of the SISO observer $\mathcal{H}(z)$ |
| $(\bar{x}, \bar{y})$ | 2-D Cartesian coordinates |
| $\sigma_{\text{tgt}}^2$ | Expected squared-distance errors at SS for a signal produced by the target process model |
| $\sigma_{\text{man}}^2$ | Expected squared-distance errors at SS for a target executing a sustained constant-g (circular) turn |
| $\sigma_{\text{sns}}^2$ | Variance of measurement noise (actual); $\sigma_{\text{sns}}^2 = \sigma_R^2$ in the absence of model mismatch |
| $\varepsilon_{\bar{x}}$ | Position estimation error in the $\bar{x}$ dimension |
| $\varepsilon_{\bar{y}}$ | Position estimation error in the $\bar{y}$ dimension |
| $\varepsilon_R$ | Radial estimation error for a target on a circular orbit |
| $\varepsilon_\theta$ | Angular estimation error for a target on a circular orbit |
| $N_{\text{frm}}$ | Number of frames used in a MC simulation scenario |
| $N_{\text{rep}}$ | Number of repetitions of a given MC simulation scenario |
| $\tilde{\sigma}_D$ | Root-mean-squared distance error over all frames and all repetitions |
| $\boldsymbol{g}_{[\cdot]}$ (or $g_{[\cdot]}$) | A vector (or element) within the $\boldsymbol{G}$ matrix for the subscripted system |
| $\mathbb{T}_{[\mathbb{C}]}^{[\mathbb{B}]\leftarrow[\mathbb{A}]}$ | A coordinate transformation matrix, from coordinate system $[\mathbb{A}]$ to coordinate system $[\mathbb{B}]$, for LSS system $[\mathbb{C}]$ |
| $\mathcal{O}_{[\mathbb{C}]}^{[\mathbb{B}]}$ | Observability matrix of LSS system $[\mathbb{C}]$ in coordinate system $[\mathbb{B}]$ |
| $[\cdot]^{\text{kin}}$ | A matrix or vector in kinematic coordinates |
| $[\cdot]^{\text{pcf}}$ | A matrix or vector in a coordinate system that reduces the LSS system to process canonical form |
| $[\cdot]^{\text{ocf}}$ | A matrix or vector in a coordinate system that reduces the LSS system to observer canonical form |